\documentclass[conference]{IEEEtran}
\IEEEoverridecommandlockouts
\usepackage{cite}

\usepackage{amsmath,amssymb,amsfonts}
\usepackage{pifont}
\usepackage{algorithmic}
\usepackage{graphicx}
\usepackage{textcomp}
\usepackage[table,xcdraw]{xcolor}
\def\BibTeX{{\rm B\kern-.05em{\sc i\kern-.025em b}\kern-.08em
    T\kern-.1667em\lower.7ex\hbox{E}\kern-.125emX}}

\usepackage{balance}       
\usepackage{graphics}      
\usepackage[T1]{fontenc}   
\usepackage{txfonts}
\usepackage{mathptmx}
\usepackage{color}
\usepackage{booktabs}
\usepackage{algorithm2e}
\RestyleAlgo{ruled}

\usepackage{caption}
\usepackage{subcaption}
\usepackage{multirow}
\renewcommand{\arraystretch}{2}

\usepackage{microtype}        
\usepackage{ccicons}          

\usepackage{todonotes}
\usepackage{mdframed}
\usepackage{svg}
\usepackage{dblfloatfix} 
\usepackage{booktabs, makecell} 

\usepackage{hyperref}
\makeatletter
\def\url@leostyle{%
  \@ifundefined{selectfont}{
    \def\UrlFont{\sf}
  }{
    \def\UrlFont{\small\bf\ttfamily}
  }}
\makeatother
\newcommand{\bpstart}[1]{\vspace{4px}\noindent{\textbf{#1}}}

\usepackage{xargs}      
\usepackage{soul}       
\usepackage{color}      
\usepackage{xspace}     
\usepackage{xpunctuate} 
\usepackage{circledsteps}  

\newcommand{\sys}{\textcolor{black}{\mbox{\textsc{WhatsNext}}}}

\newcommand{\eg}{{e.g.,}\xspace}
\newcommand{\etal}{{et~al\xperiod}\xspace}

\definecolor{lightpink}{RGB}{237,157,202}
\definecolor{lightred}{RGB}{210,121,121}
\definecolor{lightorange}{RGB}{230,170,50}
\definecolor{lightgold}{RGB}{210,194,121}
\definecolor{lightgreen}{RGB}{121,210,121}
\definecolor{lightaqua}{RGB}{121,206,210}
\definecolor{lightBlue}{RGB}{121,124,210}
\definecolor{lightpurple}{RGB}{153,102,255}
\definecolor{red}{RGB}{178,34,34}
\definecolor{gray}{RGB}{166,166,166}

\newcommandx{\guest}[3][1=]
    {\setulcolor{lightorange}{\ul{#1}} \textcolor{lightorange} 
    {[\textbf{#2:} #3]}}
\newcommandx{\jane}[2][1=] 
    {\setulcolor{lightgreen}{\ul{#1}} \textcolor{lightgreen}   
    {[\textbf{Jane:} #2]}}
\newcommandx{\shunan}[2][1=] 
    {\setulcolor{magenta}{\ul{#1}} \textcolor{magenta}   
    {[\textbf{Shunan:} #2]}}

\newcommand{\revise}[1]{\textcolor{black}{#1}}

\definecolor{visLabelOrange}{RGB}{230,159,0}
\newcommand{\visLabel}[1]{\Circled[inner color=white, fill color=visLabelOrange, outer color=visLabelOrange]{\scriptsize\textbf{#1}}}

\definecolor{stepLabelGreen}{RGB}{0,158,115}
\newcommand{\stepLabel}[1]{\Circled[inner color=white, fill color=stepLabelGreen, outer color=stepLabelGreen]{\scriptsize\textbf{#1}}}

\definecolor{userLabelBlue}{RGB}{86,180,233}
\newcommand{\userLabel}[1]{\Circled[inner color=white, fill color=userLabelBlue, outer color=userLabelBlue]{\scriptsize\textbf{#1}}}

\newcommand{\questionRetrieval}{\textcolor{userLabelBlue}{\small\textsf{Question Retrieval}}}
\newcommand{\cellRendering}{\textcolor{userLabelBlue}{\small\textsf{Cell Rendering}}}
\newcommand{\analysisThreadVisualization}{\textcolor{userLabelBlue}{\small\textsf{Analysis Thread Visualization Update}}}
\newcommand{\userInteraction}{\textcolor{stepLabelGreen}{\small\textsf{User Interaction}}}

\definecolor{treeNodeBlue}{RGB}{46,117,182}

\definecolor{linkColor}{RGB}{6,125,233}
\hypersetup{
  colorlinks,
  citecolor=black,
  filecolor=blue,
  linkcolor=blue,
  urlcolor=linkColor,
  breaklinks=true,
  hypertexnames=false,
}

\begin{document}

\title{\sys: Guidance-enriched Exploratory Data Analysis with Interactive, Low-Code Notebooks\vspace{-5mm}
\thanks{The work was done when Chen Chen\IEEEauthorrefmark{1} was an intern at Adobe Research.}
}

\newcommand{\ssp}{,\ }

\author{\IEEEauthorblockN{Chen Chen\IEEEauthorrefmark{1}\ssp
Jane Hoffswell\IEEEauthorrefmark{2}\ssp Shunan Guo\IEEEauthorrefmark{3}\ssp Ryan Rossi\IEEEauthorrefmark{3}\ssp Yeuk-Yin Chan\IEEEauthorrefmark{3}\ssp Fan Du\IEEEauthorrefmark{3}\ssp Eunyee Koh\IEEEauthorrefmark{3}\ssp Zhicheng Liu\IEEEauthorrefmark{1}}
\IEEEauthorrefmark{1}University of Maryland, College Park, MD, USA\\
\IEEEauthorrefmark{2}Adobe Research, Seattle, WA, USA\ \ \ 
\IEEEauthorrefmark{3}Adobe Research, San Jose, CA, USA\\
\IEEEauthorrefmark{1}\{cchen24, leozcliu\}@umd.edu, 
\IEEEauthorrefmark{2}\IEEEauthorrefmark{3}\{jhoffs, sguo, ryrossi, ychan, fdu, eunyee\}@adobe.com
\vspace{-3mm}
}

\maketitle



\maketitle

\begin{abstract}

Computational notebooks such as Jupyter are popular for exploratory data analysis and insight finding. Despite the module-based structure, notebooks visually appear as a single thread of interleaved cells containing text, code, visualizations, and tables, which can be unorganized and obscure users' data analysis workflow. Furthermore, users with limited coding expertise may struggle to quickly engage in the analysis process. In this work, we design and implement an interactive notebook framework, \sys, with the goal of supporting low-code visual data exploration with insight-based user guidance. In particular, we (1)~re-design a standard notebook cell to include a recommendation panel that suggests possible next-step exploration questions or analysis actions to take, and (2)~create an interactive, dynamic tree visualization that reflects the analytic dependencies between notebook cells to make it easy for users to see the structure of the data exploration threads and trace back to previous steps.

\end{abstract}


\section{Introduction}
Computational notebooks, such as Jupyter Notebook~\cite{kluyver2016jupyter}~and RStudio~\cite{racine2012rstudio}, are the most popular tools for exploratory data analysis~(EDA) among data scientists~\cite{lau2020design,li2021edassistant}.
Computational notebooks have several advantages: 
(1)~users can combine code, text, visualizations, and tables in one environment~\cite{wenskovitch2019albireo,li2021edassistant}, 
(2)~users can easily change the code to see the intermediate results and debug the behavior, and 
(3)~users can leverage the notebook framework to deploy and share notebooks with the cloud, thereby facilitating collaboration between developers~\cite{araya2018jovial}.

However, computational notebooks' code-reliance limits~their use by inexperienced programmers like sales managers and~doctors~\cite{murallie_2021} who cannot code (or do not have time to code), despite their need for tooling that can support predictive or prescriptive analysis. 
These users often collaborate with programmers to obtain the desired results, which inevitably delays decision-making.
Although Observable~\cite{observable}
aims to alleviate the need for extensive programming expertise by providing APIs and templates for quickly rendering data in a notebook environment, users are still expected to have basic programming knowledge or learn on the fly. 
To make this form of analysis more accessible to users with varying expertise, one option is to explore a less code-dependent notebook environment; however, no-code or low-code interaction remains under-explored.

Notebooks also have some navigational \mbox{challenges~\cite{guo2012burrito,wenskovitch2019albireo}.}
A basic notebook collects cells containing code, text, tables,~and visualizations into a single, interleaved thread, which may not accurately capture the user's analysis flow or the complex hierarchy of branching analysis threads. 
For example, from a single bar chart, the user may want to explore different grouped bar charts by introducing categorical attributes~(one at a time), represented as several consecutive new notebook cells.
This linear structure cannot clearly reflect the hierarchical relationship between these new cells and the source visualization. 
The user also cannot easily restore deleted cells or recall how the removed cell contributes to the analysis hierarchy, thus creating barriers to understanding the provenance of steps during data analysis~\cite{ragan2015characterizing, head2019managing, macke2020fine}. 
This navigation issue is one of the major disadvantages of using notebooks for data exploration.

To address these challenges, we introduce \sys,~an interactive notebook environment for low-code data exploration that 
(1)~recommends insight-related follow-up questions to interactively generate new visualizations, and 
(2)~visualizes~the analysis hierarchy to help users trace the history \mbox{of diverging} analysis threads. 
\sys\ augments a standard notebook cell with a no-code interaction panel showing recommended follow-up analysis questions based on the current visualization. 
\sys\ also provides a dynamic tree visualization that shows the analytic dependencies between cells; 
users~can~thus examine the overall notebook structure, view the cell-node correspondence, open the recommendation panel of a specific cell, and restore deleted cells from the analysis thread visualization. 

In summary, we contribute the design of \sys, a novel interactive notebook framework for guidance-enriched visual data analysis, and a next-step recommendation heuristic for efficiently exploring insights from the current visualization.

\section{Related Work}

\begin{figure*}[!t]
\centering
  \includegraphics[width=\textwidth]{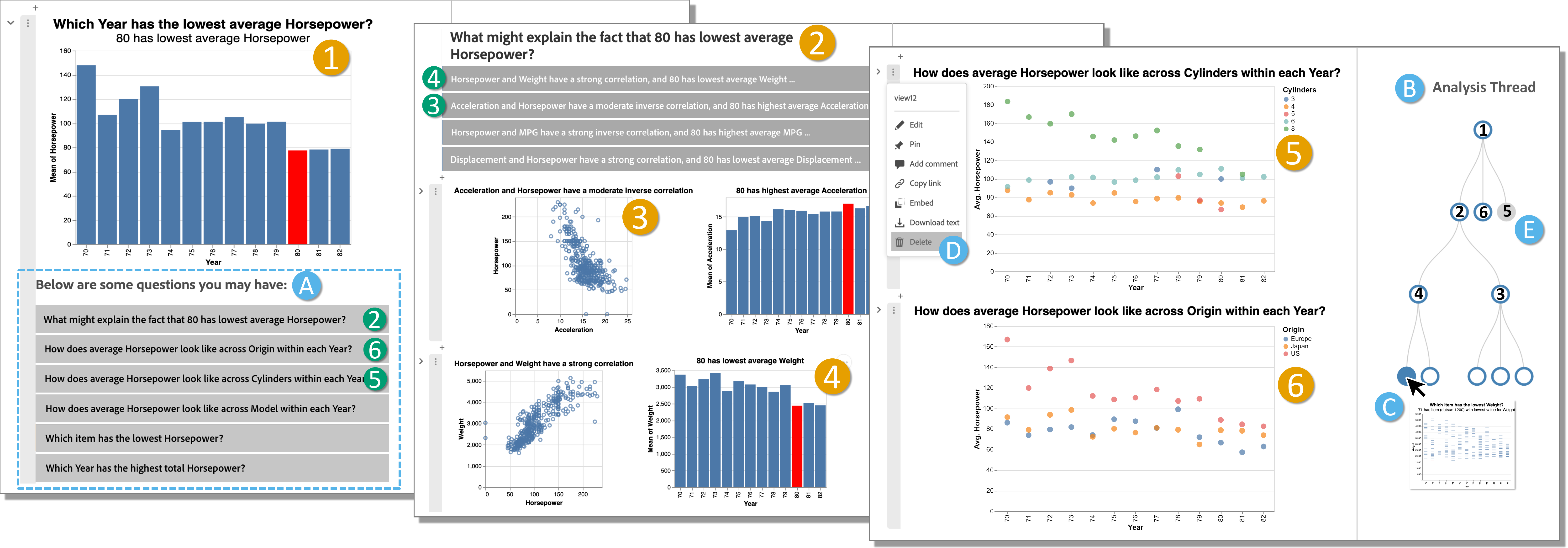}
  \caption{The interface and sample use case of \sys. The core UI features are a new, no-code interaction panel that recommends follow-up questions~\userLabel{A} and the analysis thread visualization of the exploration structure~\userLabel{B}; the user can quickly generate new notebook cells~\visLabel{2-6} by selecting recommended follow-up questions~\stepLabel{2-6}. Other simple interactions allow the user to explore, delete, and restore cells from the analysis history~\userLabel{C-E}.
  The sample cars dataset is available at \url{https://goo.gl/9G1egz}.
  }
  \label{fig:teaser}
  \vspace{-14px}
\end{figure*}

This work combines research on computational notebooks and visualization recommendation for exploratory data analysis.

\subsection{Interactions in Computational Notebook Environments}
Given their code-reliance, computational notebooks are often used for collaborative programming~\cite{wang2019data,mendez2019toward}. 
Recent work has developed helper functions or built-in widgets to facilitate the overall user experience, for example, by including semantic code search in notebook collections~\cite{li2021nbsearch} and interactive~visual exploration of search results~\cite{li2021edassistant},~\revise{enabling efficient exploration of the cells' history~\cite{kery2017variolite, kery2019towards,merino2022making, wang2022stickyland} or providing version control~\cite{kery2018interactions}, supporting quick decision-point navigation~\cite{weinman2021fork}, and detecting and resolving staleness issues~\cite{macke2020fine}. }
To facilitate code editing, 
Kery~\etal~\cite{kery2020mage,kery2020future} proposed the concept of fluidly moving between code and GUI editing;
for example, when a user repositions a table column in the output of a Jupyter notebook cell, the code will automatically update to reflect the new sort order. 
This work has also inspired research into several more interactive notebook-based prototypes, e.g.,~Unravel~\cite{shrestha2021unravel} enables structured edits via drag-and-drop and toggle interactions, and 
Symphony~\cite{bauerle2022symphony} promotes shareable, task-specific data-driven components in Jupyter for ML practitioners.
These works aim to improve the usability of notebook environments.
However, they still maintain some code-reliance and do not fully explore how to enable a low-code mode for notebooks.

\begin{figure*}[!t]
\centering
    \includegraphics[width=2\columnwidth]{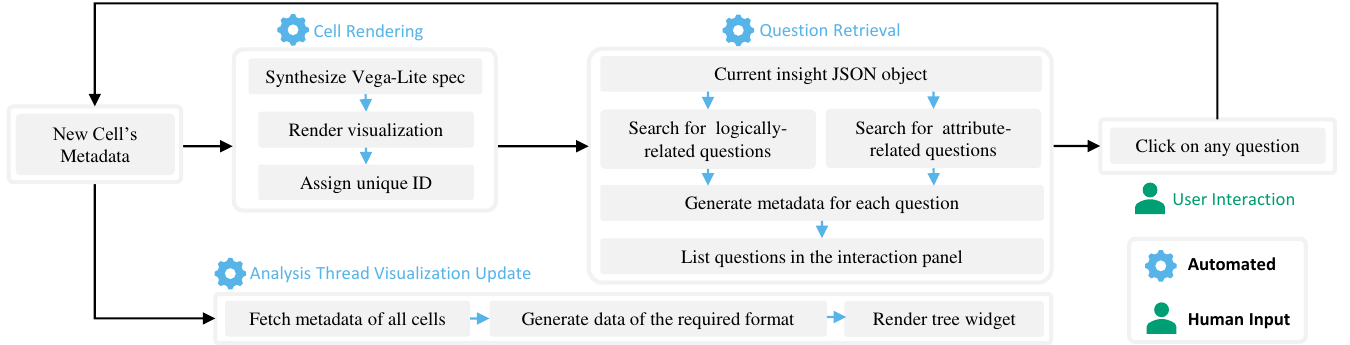}
  \caption{The architectural pipeline of \sys. With a new notebook cell's metadata, \sys\ renders the corresponding visualizations(s)~(\cellRendering), searches for both logically-related and attribute-related insights, turns them into questions, and populates these questions in the interaction panel~(\questionRetrieval). The \analysisThreadVisualization\ happens in real-time. Click events on the questions from the user~(\userInteraction) further create new cells, making the process iterative.
  }~\label{fig:system-overview}
  \vspace{-18px}
\end{figure*}

\subsection{Visualization Recommendation for EDA}
Presenting visualization recommendations can help users with data exploration and analysis~\cite{grammel2010information}. 
To better reflect user intents, many systems leverage user interactions or user inputs to refine the recommendations. 
For example, Voyager~\cite{wongsuphasawat2015voyager}~and Voyager~2~\cite{wongsuphasawat2017voyager} leverage user-specified fields and wildcards to iterate on possible data attributes, transformations, and~encodings to explore for the final visualization recommendations. 
VizAssist~\cite{bouali2016vizassist} recommends relevant visualizations based on user-specified analysis objectives~(\eg~to discover outliers~or to find correlations). 
Lux~\cite{lee2021lux} was built on Jupyter to provide real-time visualization recommendations of patterns, trends, and analysis directions with optimized computational overheads whenever the user prints a dataframe in their notebooks. 
Li~\etal~\cite{li2022structure} developed a system that recommends similar visualizations by extracting the structural information from a user-provided SVG image
with GNN-based contrastive~learning~\cite{sun2019infograph}.

With the advances in the NLP field, many systems have introduced natural language interfaces~(NLI) for visual analysis and recommendation. 
These systems can extract a user's~analytical intents by identifying explicit data attributes, numerical values, and chart types in input queries, and facilitate effective conversations with the user by modeling ambiguity properly. 
DataTone~\cite{gao2015datatone} offers interactive ambiguity widgets to let~users correct system decisions. 
Eviza~\cite{setlur2016eviza} supports language pragmatics in analytical interaction to enable conversations between a user and their data that allow follow-up queries. 
FlowSense~\cite{yu2019flowsense} uses semantic parsing and special utterances to understand the dataflow context from plain English,
facilitating the creation of multi-view linked visualizations. 
Snowy~\cite{srinivasan2021snowy} generates utterance recommendations to guide conversational visual analysis based on data interestingness and language~pragmatics.

\section{\sys: Guidance for EDA in Notebooks}

\label{sec:goals}


\sys\ explores new forms of \textit{guidance} for computational notebook environments to boost the user's awareness when performing exploratory data analysis (EDA) in a low-code manner. To this end, 
%
%
we identified four key design goals.  

\newcommand{\DGone}{\textbf{DG1: Low-code}}
\newcommand{\DGtwo}{\textbf{DG2: Insight-driven}}
\newcommand{\DGfour}{\textbf{DG3: History}}
\newcommand{\DGfive}{\textbf{DG4: Structure}}


\bpstart{\DGone.} \revise{Reduce the difficulty of data exploration~\cite{kery2017exploring}} to support users with varying levels of programming expertise. 

\bpstart{\DGtwo.} Help users better locate relevant information quickly and synthesize compound data insights~\cite{law2020characterizing}.


\bpstart{\DGfour.} 
Help users recall explored content and navigate efficiently by providing visual cues and simple~interactions~\cite{rule2018exploration}.


\bpstart{\DGfive.} 
Reveal the analytic dependencies between cells to make users aware of the overall exploration status~\cite{wang2022stickyland}.

\subsection{Interface and Pipeline}

Figure~\ref{fig:teaser} illustrates a typical use case of \sys.
To simplify exploratory data analysis, the new, no-code interaction panel~(Fig.~\ref{fig:teaser}\userLabel{A}) recommends follow-up questions or actions. Starting from a source visualization~(Fig.~\ref{fig:teaser}\visLabel{1}), the user~selects the first recommended follow-up question~(Fig.~\ref{fig:teaser}\stepLabel{2}) to automatically generate the next step in the analysis process~(Fig.~\ref{fig:teaser}\visLabel{2}). 
This new cell contains possible explanations for the selected question, which the user can select (Fig.~\ref{fig:teaser}\stepLabel{3-4}) to generate~new visualizations (Fig.~\ref{fig:teaser}\visLabel{3-4}).
To accurately capture the branching and hierarchical nature  of the analysis process, \sys\ provides an analysis thread visualization~(Fig.~\ref{fig:teaser}\userLabel{B}) to show the overall exploration structure~(nodes are annotated with the corresponding cell numbers for clarity in the paper).
Tooltips on the analysis thread visualization allow the user to preview the visualizations in the corresponding cell~(Fig.\ref{fig:teaser}~\userLabel{C}); the user can jump to any cell by clicking its node. With these interactions, the user can quickly trace back to Fig.~\ref{fig:teaser}\visLabel{1} and generate two new cells~(Fig.~\ref{fig:teaser}\visLabel{5,6}) to explore different breakdowns by year.
The user can delete unwanted cells from the cell menu~(Fig.~\ref{fig:teaser}\userLabel{D}); the analysis thread visualization maintains corresponding nodes in gray~(Fig.~\ref{fig:teaser}\userLabel{E}) as a history from which to restore old cells. 
The architectural pipeline of \sys\ is shown in Fig.~\ref{fig:system-overview}.

\subsection{The Low-code Design of Notebook Cells}
\revise{A standard notebook cell consists of two parts: a code editor containing the corresponding code snippet and the executed result~\cite{kluyver2016jupyter, racine2012rstudio}. 
To support data exploration in a low-code manner, we augment the notebook cells in \sys\ to include an interaction panel with recommendations for new steps in the analysis process in the form of follow-up questions or actions~(Fig.~\ref{fig:teaser}\stepLabel{2-6}).
Without writing any code, the user can click on the options presented in the interaction panel, which inserts new notebook cells accordingly. This interaction allows users to focus on the analysis process directly, rather than the intricacies of writing the correct code to meet their needs.}

\subsection{Question-driven Data Insight Exploration}\label{sec:4.3}
\sys\ starts with an initial data insight visualization and the underlying tabular data~(e.g., both of which can be retrieved from a dashboard), and renders the first notebook cell embedding the visualization~(Fig.~\ref{fig:system-overview}: \cellRendering). Meanwhile, the back-end utilizes algorithms from Voder~\cite{srinivasan2018augmenting} to generate a search space of insights for later steps in the pipeline. We choose Voder as the insight generator mainly due to the JSON format and useful information provided with the insights, such as: \texttt{activeHtml}~(the insight text), \texttt{attributes}~(involved~data attributes), \texttt{relatedVisObjects}~(candidate visualizations, \eg mark type, encodings, aggregation), \texttt{tier}~(the importance level), and \texttt{type}~(the insight type). Voder supports four types of insights: \emph{anomalies}, \emph{correlation}, \emph{distribution}, and \emph{extremum}. Some insights involve value derivation~(e.g.,~computing the average) and value filtering~(e.g.,~anchoring on a specific value).

\newcommand{\systemRequire}{\textbf{Running System Needed}}

\sys\ then retrieves potential follow-up questions for the visualization~(a.k.a., the insight) and displays them in the interaction panel of the cell~(Fig.~\ref{fig:system-overview}: \questionRetrieval). \sys\ considers two kinds of follow-up questions: \emph{logically-related questions}, whose answers are logically connected to the current insight, and \emph{attribute-related questions}, whose answer includes insights involving the same attributes as the current cell. For example, given the insight \textit{``Cars from the year 1980 have the lowest average horsepower''}~(Fig.~\ref{fig:teaser}\visLabel{1}),
\revise{\sys\ generates six questions, including both logically-related questions (e.g., \textit{``What might explain the fact that 1980 has the lowest average horsepower?''}) and attribute-related questions (e.g., \textit{``Which item has the lowest horsepower?''}) as options for how the user might proceed with their analysis.}

Attribute-related questions are straightforward to retrieve: the system obtains the attribute set for the current cell's insight, searches for insights whose attributes overlap, and converts them into questions. For logically-related questions, the system needs to determine which insights are logically related to the current insight; we thus propose retrieval rules based on the insight type~(Table~\ref{table:why-question-taxonomy}) to search for logically-related insights to present as questions. Table~\ref{table:why-question-taxonomy} also provides example insights.

\newcolumntype{P}[1]{>{\centering\arraybackslash}p{#1}}
\renewcommand{\arraystretch}{1}

\newcommand{\centerCell}[1]{\centering #1}

\begin{table*}[h!]
    \centering
    \caption{Rules for logically-related insight retrieval based on the insight type. \sys\ searches for an insight (combination) that has a logical connection to the given insight to support reasoning and drill-down analysis, and reflect low-level analytic tasks. Four insight types are considered: [Ext]remum, [Cor]relation, [Ano]maly, and [Dis]tribution.
    }
    \resizebox{\textwidth}{!}{
    \begin{tabular}{ m{0.25\linewidth} m{0.47\linewidth} m{0.225\linewidth} } 

\toprule

    Given Insight Example
    &
    Logically-related Insight Example
    &
    Converted Question \\

\midrule 
    
     \multirow{3}{\linewidth}[-1em]{[Ext] \textit{``Cars from the year 1980 have the \textbf{lowest} average Weight''}}
     & 
     [Ext] \textit{``Cars from the year 1980 have the \textbf{lowest} average Horsepower''} + [Cor] \textit{``Horsepower and Weight have a strong \textbf{correlation}''}
     & 
     \multirow{3}{\linewidth}[-1em]{\textit{``Why do cars from the year 1980 have the \textbf{lowest} average Weight?''}} \\
    
\cmidrule(lr){2-2} 

    & [Ano] \textit{``There are three \textbf{anomalies} regarding Weight in the year 1980''}
    &  \\
    
\cmidrule(lr){2-2}

    & [Ext] \textit{``Cars from Japan in the year 1980 have the \textbf{lowest} average weight''}
    &  \\
    
\midrule
    
    [Cor] \textit{``Horsepower and Weight have a strong \textbf{correlation}''} 
    & 
    [Cor] \textit{``Weight and Displacement have a strong \textbf{correlation}''} + {\textcolor{white}{\_\_\_\_\_\_}} [Cor] \textit{``Horsepower and Displacement have a strong \textbf{correlation}''} 
    & 
    \textit{``Why do Horsepower and Weight have a strong \textbf{correlation}?''} \\

\midrule
    
    [Ano] \textit{``The car `renault 18i' appears to be an \textbf{outlier} regarding Horsepower''} 
    & 
    [Dis] \textit{``Most values for Horsepower are in the \textbf{range} [75.0, 125.0]''} 
    & 
    \textit{``What is the major value \textbf{range} of Horsepower?''} \\

\midrule

    \multirow{2}{\linewidth}[-0.7em]{[Dis] \textit{``Most values for Horsepower are in the \textbf{range} [75.0, 125.0]''}}
    & 
    [Ano] \textit{``The car `renault 18i' seems an \textbf{outlier} regarding Horsepower''} 
    & 
    \textit{``What are potential \textbf{outliers} regarding Horsepower?''} \\

\cmidrule(lr){2-3}

    & 
    [Dis] \textit{``Most values for Horsepower in 1980 are in the \textbf{range} [70.0, 120.0]''}
    & 
    \textit{``What is the \textbf{distribution} of Horsepower in the year 1980?''} \\

\bottomrule

    \end{tabular}
    }
    \label{table:why-question-taxonomy}
    \vspace{-2px}
\end{table*}

\newcommand{\insight}[1]{\emph{#1}} 

\bpstart{Extremum.} For an \insight{extremum} insight involving a categorical variable $c1$ and a quantitative variable $q1$, \sys\ looks for three kinds of logically-related insights: (1)~an~\insight{extremum} insight involving the same categorical variable $c1$ and another quantitative variable $q2$ plus a \insight{correlation} insight between these two quantitative variables $[q1,\ q2]$, (2)~an~\insight{anomaly} insight with the same variable pair ($c1$ and~$q1$), and (3)~an \insight{extremum} insight with an additional categorical variable $c2$. All three insight types reveal possible explanations for the current \insight{extremum}~insight.

\bpstart{Correlation.} For a \insight{correlation} insight involving two quantitative variables $[q1,\  q2]$, the system identifies two \insight{correlation} insights that involve $[q1,\  q3]$ and $[q2,\  q3]$ respectively, to~expand on the current \emph{correlation} with two \mbox{additional correlations.}

\bpstart{Anomaly.} For an \insight{anomaly} insight, \sys\ looks for a \insight{distribution} insight involving the same quantitative variable $q1$ to further present the value range for the majority of the data.
    
\bpstart{Distribution.} For a \insight{distribution} insight, the system looks for (1)~an \insight{anomaly} with the same quantitative variable to show potential outliers, and (2)~a \insight{distribution} insight with an extra categorical variable to reveal drill-down distribution~statistics.

The above rules were developed based on three design~goals: (1)~support potential reasoning about a~data insight, (2)~enable drill-down analysis by introducing new attributes, and~(3)~reflect low-level analytic tasks (e.g.,~\textit{Find Anomalies} or~\textit{Correlate}~\cite{amar2005low}).
Note that these rules do not require insights to come from Voder~\cite{srinivasan2018augmenting}; insights simply need to be formatted accordingly. 

After the search process, \sys\ uses a simple but effective template-based method~\cite{fabbri2020template} to generate questions from the selected insights. Typically there are multiple questions~(generated from multiple insights) to display to the user. Thus, \sys\ must order the recommended insights appropriately; the interaction panel ordering follows two principles: (1)~logically-related questions have higher priority than attribute-related questions, and (2)~attribute-related questions are ordered according to the importance level from~Voder~\cite{srinivasan2018augmenting}. 
When the user selects a question that has multiple possible answers, \sys\ will present the text of each insight combination in the interaction panel as a possible \textit{action} to take~(Fig.~\ref{fig:teaser}\stepLabel{3,4}), which functions similarly to a \textit{question}. 

\subsection{The Analysis Thread Visualization}\label{sec:4.4}

The analysis thread visualization is updated in real-time to show the structure of the EDA process when notebook cells are \emph{added} or \emph{deleted}~\mbox{(Fig.~\ref{fig:system-overview}:~\textcolor{lightBlue}{\analysisThreadVisualization})}.
When deleting a cell that the user deems unnecessary~(Fig.~\ref{fig:teaser}\userLabel{D}), 
the analysis thread visualization will grey out the corresponding node to indicate its archival status~(Fig.~\ref{fig:teaser}\userLabel{E}) and maintain~its child nodes, if any. The user can restore deleted cells by simply clicking its corresponding node and selecting ``restore this cell''. Two additional node interactions are included to support user navigation: (1)~\emph{hover} shows a mini version of the visualization in the corresponding cell as a tooltip, and (2)~\emph{click} jumps the user directly to the corresponding cell in the notebook. 
The analysis thread visualization captures the analytic dependencies between notebook cells that would otherwise be hidden or lost in a linear notebook; this view can thus act as a more complete archive of the user's overall EDA process for quick review.

\section{Discussion, Limitations, and Future Work}

We discuss the system limitations and three areas for future work to enhance the practical quality of~\sys.

\bpstart{(1) Improve connections between the notebook and analysis structure.} 
The analysis thread visualization shows the hidden analytic dependencies between the linearly-ordered cells in the notebook; however, these connections can be further enhanced to reveal other hidden features (e.g., adding 
attribute or insight-type tags on recommendations or color-encoding nodes to show the recommendation type or unexplored EDA directions).

\bpstart{(2) Support user-customized EDA trajectories.} 
\sys\ recommends a variety of analysis directions, but these currently cannot be further customized by the user.
For example, users might want to specify their analysis intent explicitly or refine the follow-up questions.
Future work should determine how best to support these goals while maintaining 
ecological synchronization between the cells and analysis thread~visualization. 

\bpstart{(3)~Support thread decomposition, switching, and saving.} 
\sys\ currently supports multi-thread analysis within~a single notebook. While the analysis thread visualization can show the overall analysis structure, (1)~the~tree may get too large to effectively support navigation, and (2)~the user may want to decompose their exploration into disjoint threads. 
Future work could allow users to 
easily switch amongst analysis sub-threads to reduce the mental difficulty when tracing complex analysis histories.
\sys\ could also allow users to generate and export data insight reports to  share their analysis results.

\section{Conclusion}
\sys\ contributes an interactive notebook framework for low-code visual exploratory data analysis that recommends next-step analysis questions based on the visualization attributes and insights. To support navigation, \sys\ visualizes the analytic dependencies of diverging analysis trajectories.

\balance{}

\bibliographystyle{IEEEtran}
\bibliography{references}

\end{document}